\documentclass[twocolumn,trackchanges]{aastex631}

\usepackage{orcidlink}
\usepackage{threeparttable}
\usepackage{mathcomp} 
\usepackage{needspace}
\usepackage{bm} 

\begin{document}

\title{Turbulent Magnetic Fields and Molecular Cloud Interactions in the Supernova Remnant G1.9+0.3}

\author{Moeri Tao \orcidlink{0009-0008-9030-760X}}
\affiliation{School of Advanced Science and Engineering, Wseda University, 3-4-1 Okubo, Shinjuku-ku, Tokyo, Japan}

\author{Takaaki Tanaka \orcidlink{0000-0002-4383-0368}}
\affiliation{Department of Physics, Konan University, 8-9-1 Okamoto, Higashinada, Kobe, Hyogo, Japan}

\author{Hidetoshi Sano \orcidlink{0000-0003-2062-5692}}
\affiliation{Faculty of Engineering, Gifu University, 1-1 Yanagido, Gifu 501-1193, Japan}

\author{Rami Z. E. Alsaberi \orcidlink{0000-0001-5609-7372}}
\affiliation{Faculty of Engineering, Gifu University, 1-1 Yanagido, Gifu 501-1193, Japan}
\affiliation{Western Sydney University, Locked Bag 1797, Penrith NSW 2751, Australia}

\author{Jun Kataoka \orcidlink{0000-0003-2819-6415}}
\affiliation{School of Advanced Science and Engineering, Wseda University, 3-4-1 Okubo, Shinjuku-ku, Tokyo, Japan}

\begin{abstract}
We present on results of a spatially resolved spectral analysis of G1.9+0.3, the youngest known supernova remnant in the Galaxy. The X-ray spectra are well described by synchrotron emission from a power-law electron distribution with an exponential cutoff. We found a cutoff energy $\epsilon_0 \sim 1 ~ \rm{keV}$ in both the radio bright rim and the X-ray bright rims. In the loss-limited case, the cutoff energy depends on the shock velocity $v_{\rm{sh}}$ and the Bohm factor $\eta$, following the relation $\epsilon_0 \propto v_{\rm{sh}}^2 \eta^{-1} $. Our analysis shows that $\eta$ ranges from 2 to 4 in the radio rim and from 12 to 15 in the X-ray rims. This suggests that the magnetic field in the radio rim is more turbulent than in the X-ray rims. The presence of CO clouds along the radio rim likely contributes to this difference. Interaction between the shock and these clouds can slow the shock down and generate turbulent eddies. The resulting turbulence eddies can amplify the magnetic field. We propose that the strong radio emission from the radio rim is primarily due to  this amplified magnetic field. In contrast, a CO cloud located in the south-west appears to lie in the foreground, as indicated by its low turbulence and the absence of shock deceleration.

\end{abstract}


\section{Introduction} \label{sec:intro}
Young supernova remnants (SNRs) are considered promising candidates for accelerating Galactic cosmic rays through diffusive shock acceleration (DSA). However, the detailed mechanisms of DSA are still not fully understood. In this process, particles gain energy by repeatedly crossing the shock front in both forward and backward \citep{drury1983introduction}. The acceleration of particles is limited either by the age of the remnant, energy losses due to radiation, or escape. In the synchrotron-loss-limited case, the acceleration timescale equals the synchrotron cooling timescale, which is shorter than the age of the remnant. Assuming Bohm diffusion, the maximum energies of accelerated particles in the age-limited case and loss-limited case are $E_{\rm{max}}(\rm{age}) \propto \eta^{-1}$$Bv_{\rm{sh}}^2$ and $E_{\rm{max}}(\rm{loss}) \propto \eta^{-1/2}$$B^{1/2}v_{\rm{sh}}$, respectively. $\eta$ is the Bohm factor, the ratio of the particle’s mean free path to its gyroradius. The Bohm factor $\eta$ characterizes the level of magnetic field turbulence and the efficiency of particle acceleration. In case of Bohm limit ($\eta = 1$) , the magnetic field is highly turbulent, and acceleration is maximally efficient. Therefore, understanding the Bohm factor and the magnetic field strength is essential for probing particle acceleration mechanisms in young SNRs.

\begin{figure*}[t]
\begin{center}
    \includegraphics[scale=0.26]{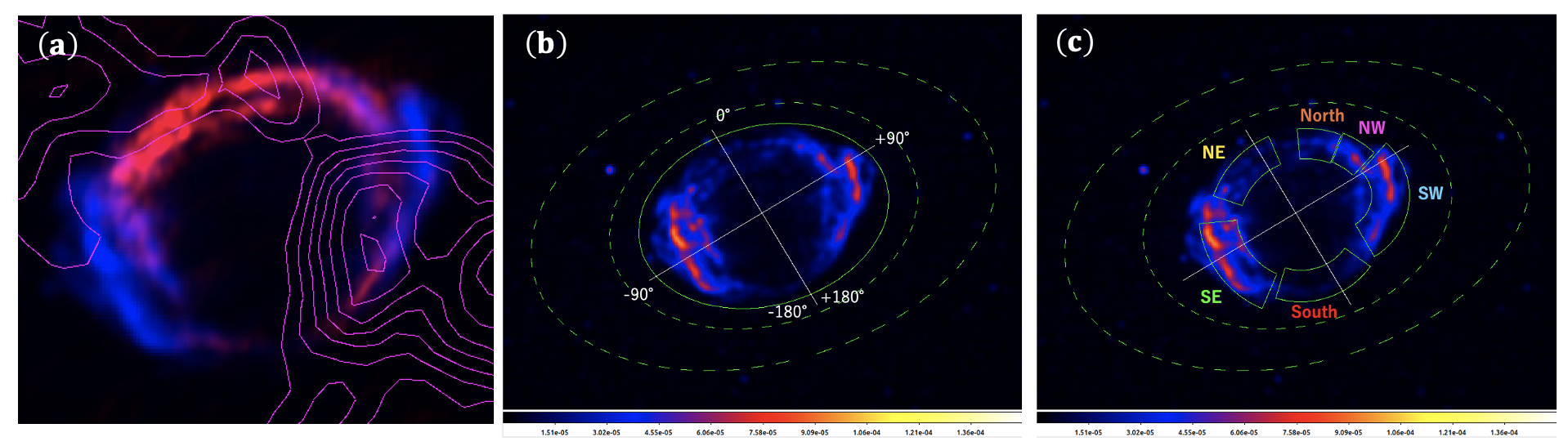}
    \caption{(a) Two-color image of G1.9+0.3: $Chandra$ X-ray data is shown in blue, and the 9 GHz radio data (obtained with ATCA) is shown in red \citep{enokiya2023discovery}. The overlaid contour represents the integrated intensity distribution of the associated cloud in $^{12}$CO($J$~=~3--2). The contour levels of CO are 25.000, 28.125, 31.250, 34.375, 37.500, 40.625, 43.750, 46.875, and 50.000 K~km~s$^{-1}$. (b) Exposure-corrected $Chandra$ X-ray image (0.5$-$7.0 keV) from 2011 observations. The solid green line marks the source region (entire remnant), and the dotted green line indicates the background region. (c) Same as (b), but showing source regions for NE, North, NW, SW, South, and SE sectors (outlined in solid green). The color scale represents flux in units of $\rm{count~ s^{-1}}$. Both images are smoothed using a Gaussian kernel with $\sigma = 1''5$. White crosshairs indicate reference axes for the azimuthal profile shown in Figure \ref{fig:angle_vs_eta}, aligned with Galactic longitude and latitude.}
    \label{fig:region}
    \end{center}
\end{figure*}

G1.9+0.3 is the youngest known Galactic SNR with an estimated age of approximately 100 years \citep{reynolds2008youngest}. One of its most striking features is the difference in radio and X-ray morphologies (see Figure \ref{fig:region} (a)). The X-ray emission, observed with Chandra (shown in red), displays a bilateral symmetry similar to that of SN 1006. In contrast, the radio emission, observed with the Australia Telescope Compact Array (ATCA; shown in blue), exhibits a single maximum located north of the remnant. \cite{brose2019nonthermal} proposed a two-shock model to explain this difference, in which the X-ray and radio emissions are primarily produced by the forward and reverse shocks, respectively.

Previous studies have measured the expansion speed of G1.9+0.3. Notably, \cite{borkowski2017asymmetric} found a variation in expansion velocity by a factor of five, despite the remnant's young age. The slowest shocks, with the speed around $3600 ~\rm{km~s^{-1}}$, occer in the radio bright rims (hereafter reffered to as radio rim). In contrast, the X-ray bright rims (X-ray rims) exhibit much faster shock, with the velocities of $ 12,000-13,000 ~\rm{km~s^{-1}}$. \cite{borkowski2017asymmetric} suggested that the slower expansion in the radio rim results from the shock interacting with denser ambient material. Indeed, \cite{enokiya2023discovery} reported that the presence of molecular cloud associated with G1.9+0.3. This CO cloud exhibits three peaks located in the north, north-east (NE), and south-west (SW). The northen and NE peaks spatially coincide with the radio rim, while the SW peak corresponds to the X-ray rim.

X-ray observations by $Chandra$ and $NuSTAR$ \citep{reynolds2009x, zoglauer2014hard, aharonian2017energy} have shown that the spectrum of G1.9+0.3 can be well modeled as synchrotron emission originating from a power-law distribution of electrons with an exponential cutoff. Previous studies reported characteristic cutoff energies of approximately 2 keV using the $srcut$ model \citep{reynolds2008youngest} and around 1.5 keV when using a power-law model with an exponential cutoff.

In this study, we conducted a spatially resolved spectral analysis that considers the distribution of molecular clouds, in order to explain the difference in radio and X-ray morphologies. To investigate the relationship between these morphologies and the molecular cloud distribution, we divided the remnant into six regions: north-east (NE), north, north-west (NW), south-west (SW), south, and south-east (SE). The data analysis and corresponding results are presented in Section \ref{sec:analysis}. In Section \ref{sec:discussion}, we discuss the relationship among the Bohm factor $\eta$, the CO cloud distribution, and the observed radio flux. Finally, our conclusions are summarized in Section \ref{sec:summery}.

\needspace{5\baselineskip}

\section{Data Analysis and Result} 
\label{sec:analysis}
$Chandra$ observed G1.9+0.3 in May and July 2011. For spectral fitting, five long observations (IDs : 12689, 12691, 12692, 12693, and 12694) were combined to improve photon statistics, with a combined effective exposure time of 787 ks. To avoid the effects of expansion and brightening, only data from 2011 were used, following the approach of \cite{borkowski2013supernova} and \cite{zoglauer2014hard}. Data reduction and spectral extraction were carried out using the Chandra Interactive Analysis of Observations (CIAO) software version 4.17 and the Chandra Calibration Database (CALDB) version 4.12.0.


\begin{table} [h]
     \centering
    \caption{Best-fit Spectral Parameters}
    \begin{tabular}{c|ccc}
     \hline
    $~$ Region $~$ & $N_{\rm{H}}$ ($\times 10^{22} \rm{cm}^{-2}$)& $~$$\epsilon_0$ (keV) $~$&$\chi^2$/dof\\
     \hline
     Entire & $7.14^{+0.09}_{-0.09}$ & $1.13^{+0.14}_{-0.11}$ & $2318.36/2151$\\
     NE  & $7.00^{+0.35}_{-0.33}$ & $0.89^{+0.45}_{-0.23}$ & $377.38/409$\\
     North & $7.00^{+0.43}_{-0.40}$ & $0.72^{+0.40}_{-0.21}$ & $233.42/256$\\
     NW & $6.87^{+0.36}_{-0.33}$ & $1.44^{+1.34}_{-0.53}$ & $342.86/318$\\
     SW & $7.82^{+0.20}_{-0.19}$ & $1.06^{+0.27}_{-0.19}$ & $1052.16/1046$\\
     South & $6.81^{+0.58}_{-0.53}$ & $0.63^{+0.48}_{-0.22}$ & $252.40/253$\\
     SE & $7.29^{+0.14}_{-0.14}$ & $1.31^{+0.29}_{-0.21}$ & $1361.20/1378$\\
     \hline
    \end{tabular}
    \label{tab:fitingresult}
    \begin{flushleft}
    \footnotesize
     Note: The error refer to the 90 \% confidence level.
    \end{flushleft}
\end{table}

We divided G1.9+0.3 into six regions for spatially resolved spectral analysis: NE, North, NW, SW, South, and SE (see Figure \ref{fig:region} (c)). CO clouds are located in the north, NE, and SW regions \citep{enokiya2023discovery}. The SW and SE regions correspond to X-ray rims, while the NE and north regions correspond to radio rim. The southern region is relatively faint in both radio and X-rays. The point
sources within the background region were excluded.  Following \cite{reynolds2009x}, we adopted abundances from \cite{grevesse1998standard}. Spectral analysis was performed in the 0.5$-$8.0 keV energy band using XSPEC version 12.13.0.

We applied the synchrotron radiation model for cooling-limited electrons proposed by \cite{zirakashvili2007analytical} (hereafter referred to as the ZA07 model). This model provides an analytical solution for the synchrotron photon spectrum in the downstream region is following form \citep{zirakashvili2007analytical}
 
\begin{equation}
    \frac{dN}{d\epsilon} \propto \left(\frac{\epsilon}{\epsilon_0}\right)^{-2} \left[1+0.38\left(\frac{\epsilon}{\epsilon_0}\right)^{-\frac{1}{2}}\right]^{\frac{11}{4}} \rm{exp}\left[-\left(\frac{\epsilon}{\epsilon_0}\right)^{-\frac{1}{2}}\right],
    \label{eq:Zira}
\end{equation}

with $\epsilon_0$ is the cutoff parameter. In this model, we adopted the ratio of the upstream and downstream magnetic field, $\kappa = 1/\sqrt{11}$. 
The results of the spectral fitting using this model are summarized in Table 1. We note that a high column density was derived, consistent with values reported in previous studies ($N_{\rm{H}}=7.23^{+0.07}_{-0.07}\times 10^{22} \rm{cm}^{-2}$). The cutoff energy in the entire region $\epsilon_0 = 1.13^{+0.13}_{-0.11}~\mathrm{keV}$ is almost consistent with $\epsilon_0 = 1.1-1.2~ \rm{keV} $ obtained by \cite{tsuji2021systematic} using the ZA07 model and $\epsilon_0 \sim 1.5 ~\rm{keV}$ obtained by\cite{aharonian2017energy} with a power-law model with an exponential cutoff model. While \cite{tsuji2021systematic} and \cite{aharonian2017energy} utilized both $\bm{Chandra}$ and  $NuSTAR$ data, we use only $Chandra$ data for the detailed spatially resolved spectral analysis, which requires its excellent angular resolution. We also note that our spectral extraction region differs from theirs.


\section{Discussion} \label{sec:discussion}
\subsection{Estimation of the Bohm factor}
In Section \ref{sec:analysis}, we present spatially resolved spectral analysis using the ZA07 model. This model assumes that the energy loss of electrons is dominated by synchrotron cooling. Hence, its validity must be tested. For an equipartition between the electron and magnetic field energy densities ($U_{\rm{e}}=U_{\rm{B}}$), the minimum magnetic field strength is

\begin{equation}
     B_{\rm min}~[\mu \rm G] = 27\Bigl( {\frac{\xi \hspace{1mm} {\it d}[\rm kpc]^2 {\it f} [\rm Jy]}{{\it V} [\rm pc^3]} \Bigr)^{2/7}},
 \end{equation}
where $\xi$ is the ratio of the energy stored in electrons and protons, $d$ and $V$ are the distance and volume of the SNR, respectively, and $f$ is the flux density at $1~\rm{GHz}$ (e.g., \citeauthor{Longair}~\citeyear{Longair}). For $\xi \sim 100$ (a typical CR composition), the magnetic field $B_{\rm{min}}$ is estimated to be $\sim 100 ~ \mu \rm{G}$. The synchrotron cooling timescale $t_{\rm{synch}}$ can be expressed as
\begin{equation}
    t_{\rm{synch}} ~[\rm{yr}]\simeq 50 \left( \frac{B}{100 \mu \rm{G}}\right)^{-\frac{3}{2}}\left( \frac{\epsilon_0}{1 \rm{keV}}\right)^{-\frac{1}{2}}.
\end{equation}
In the loss-limited case, this timescale must be less than the age of the remnant ($t_{\rm{synch}} \leq t_{\rm{age}}$ ). For $t_{\rm{age}} = 110~\rm{yr}$ \citep{carlton2011expansion} and $\epsilon_0 \sim 1~\rm{keV}$, the magnetic field should be larger than $60~\mu \rm{G}$. This is acceptable for G1.9+0.3.

\begin{figure}[t]
    \centering
    \includegraphics[width=1.0\linewidth]{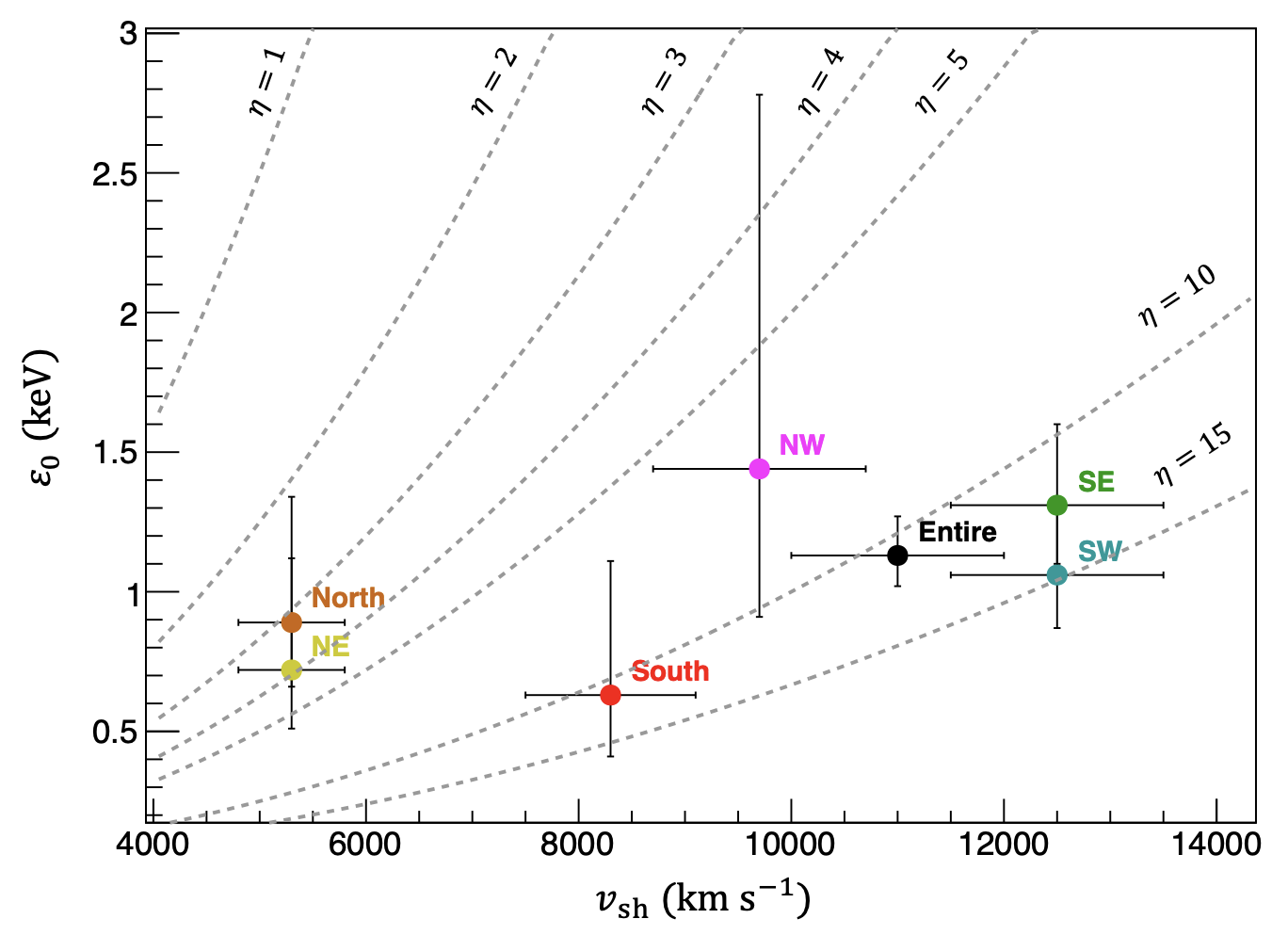}
    \caption{Relationship between the shock velocity and cutoff energy with equi-$\eta$ curves: $\eta = 1,~2,~3,~4,~5,~10,~ \rm{and} ~15$ (Equation \ref{eq:eps_v_eta}).}
    \label{fig:v_vs_ep}
\end{figure}

According to \cite{zirakashvili2007analytical}, the cutoff energy in the synchrotron spectrum and the shock speed have the following relationship;

\begin{equation}
    \epsilon_0 ~[\rm{keV}]= 2.86 {\left( \frac{\mathit{v}_{\rm{sh}}}{5300 ~\rm{km~ s^{-1}}}\right)}^2 \eta^{-1},
    \label{eq:eps_v_eta}
\end{equation}

where $v_{\rm{sh}}$ is the shock speed and $\eta$ is the Bohm factor. This equation assumes $\kappa = 1/\sqrt{11}$, same as in Equation \ref{eq:Zira}. By combining the shock velocities reported in \cite{borkowski2017asymmetric} with the cutoff energies obtained from spectral fitting using the ZA07 model, we estimate the Bohm factor $\eta$. The proper motion measurements by \cite{borkowski2017asymmetric} is based on deep Chandra observations and employs the nonparametric "Demons" algorithm, which can account for complex motions. Although the paper does not explicitly mention the projection effect, it is expected to be less significant in the outer regions, such as the rims. The relationship between shock velocity and cutoff energy for each region is shown in Figure \ref{fig:v_vs_ep}. The dotted curves represent the theoretical relations (Equation \ref{eq:eps_v_eta}) for various values of $\eta$. In the X-ray rims (SE and SW), where the shock velocity is about 
$12,500~\mathrm{km ~s^{-1}}$, $\eta$ is estimated to be in the range of 12-15. This is consistent with the values reported by \cite{tsuji2021systematic}, who found $\eta = 12-15$. In contrast, smaller $\eta$ values were estimated for the radio rim (NE and north), where the shock velocity is approximately $ 5300~\rm{km~ s^{-1}}$. We found $\eta = 2-4$, which is roughly consistent with \cite{tsuji2021systematic} ($\eta = 1-2$). 

The Bohm factor characterizes magnetic field turbulence and is defined as $\eta = (B/\delta B)^2$, where $B$ is the magnetic field strength and $\delta B$ is the amplitude of magnetic fluctuations. Therefore, a smaller $\eta$ indicates a more turbulent magnetic field. These results suggest that the magnetic field in the radio rim is significantly more turbulent than in the X-ray rims, as the $\eta$ values in the radio rims are nearly four times smaller.

\subsection{Shock-cloud interaction}
\begin{figure}[h]
    \centering
    \includegraphics[width=1.05\linewidth]{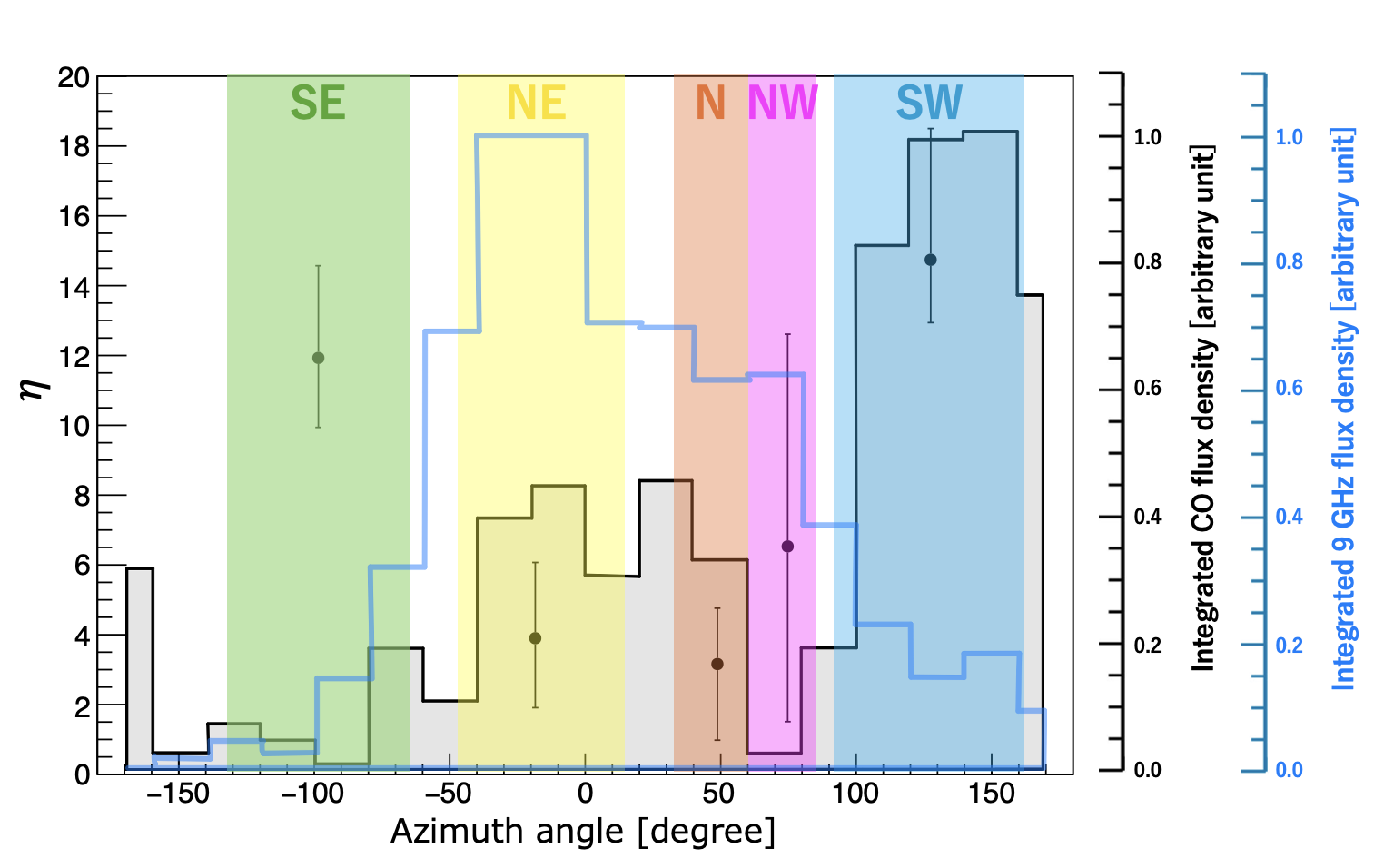}
    \caption{Azimuthal profiles of G1.9+0.3, centered at $(l,b)=(1.8710^\circ,~ 0.3237^\circ)$ (see Figure \ref{fig:region}). Black points represent the spatial variation of the Bohm factor $\eta$. The gray shaded area shows the integrated $^{12}$CO($J$=3-2) flux density, while the region marked by the blue line shows the integrated 9 GHz radio flux density (based on Figure 9 from \citeauthor{enokiya2023discovery} ~\citeyear{enokiya2023discovery}})
    \label{fig:angle_vs_eta}
\end{figure}

Figure \ref{fig:angle_vs_eta} shows the spatial distribution of the Bohm factor $\eta$, compared with that of CO clouds and radio flux. The integrated CO flux density and 9 GHz radio flux density were obtained from \cite{enokiya2023discovery}. The profile was centered at $(l,b)=(1.8710^\circ, 0.3237^\circ)$, and the direction perpendicular to the Galactic latitude was set to $0^\circ$ as shown in Figure 9 of \cite{enokiya2023discovery}. As shown in Figure \ref{fig:angle_vs_eta}, the NE and north regions, characterized by lower shock velocities and smaller $\eta$, coincide well with the CO cloud peaks. This suggests that magnetic field turbulence and shock deceleration are induced by shock-cloud interactions. Furthermore, \cite{enokiya2023discovery} found a cavity-like structure in a position-velocity diagram of $^{12}$CO($J$~=~3--2) emission (see Fig.8 of \citeauthor{enokiya2023discovery}~\citeyear{enokiya2023discovery}). The spatial extent of the CO cavity is roughly consistent with that of the radio continuum shell. This provides further support for shock-cloud interaction because such a cavity-like structure indicates an expanding gas motion and it is thought to be formed by strong stellar winds from massive progenitor of the SNR (e.g., \citeauthor{koo1990detection}~\citeyear{koo1990detection}; \citeauthor{koo1991survey}
~\citeyear{koo1991survey}). 

In the CO peak region in the SW, considering that $\eta$ is relatively high and the shock is not stalled, as is also the case in the SE, it appears that shock-cloud interaction has not occurred there. Additionally, the column density $N_{\rm{H}}$ in the SW is higher than that in the other regions, as shown in Table \ref{tab:fitingresult}. This implies that the CO cloud in the SW just lies foreground.

\needspace{5\baselineskip}
\subsection{Implication from SED }
\begin{figure}[t]
    \centering
    \includegraphics[width=1.0\linewidth]{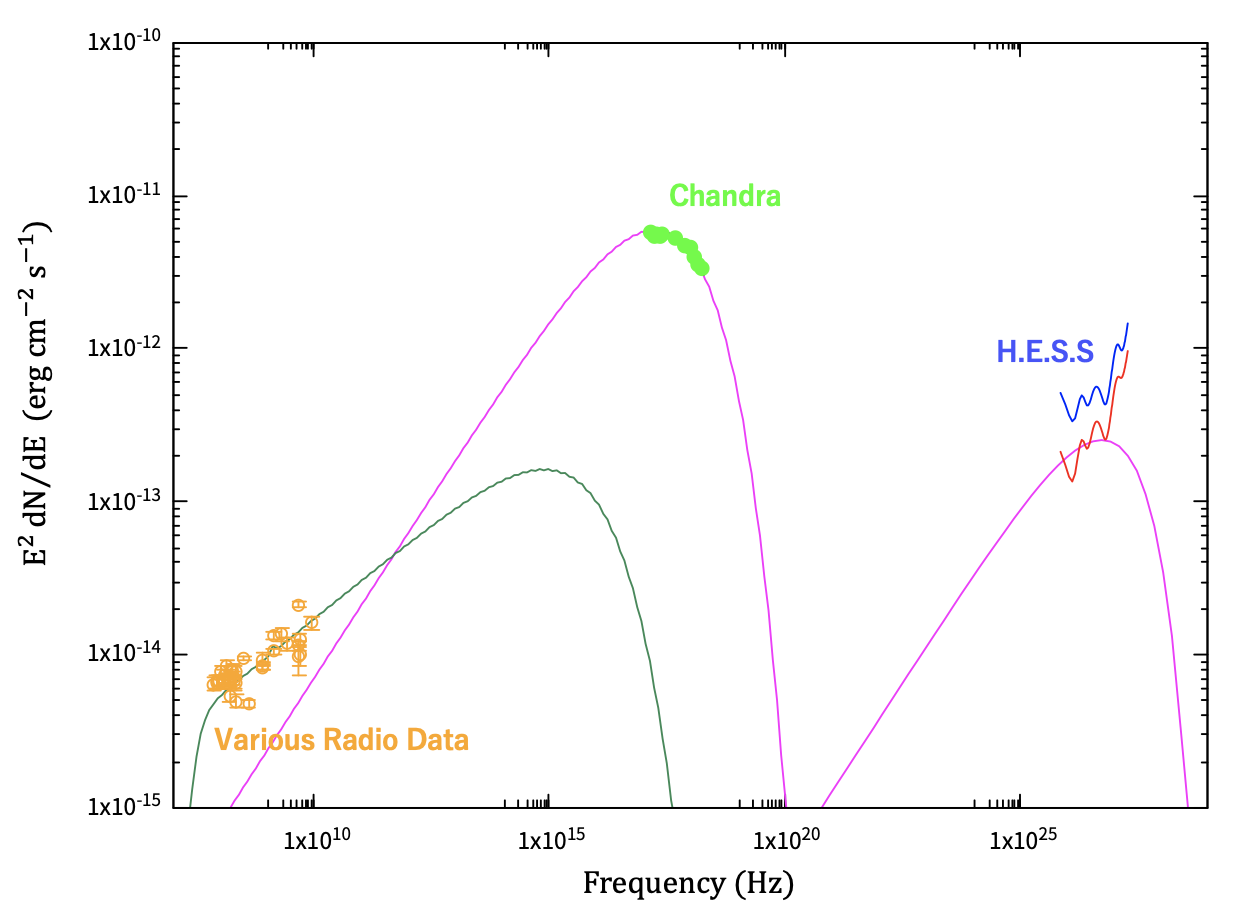}
    \caption{Spectral energy distribution of G1.9+0.3 as a whole. Radio data (orange open circles) were compiled from \cite{green2008radio} and \cite{luken2020radio}. The H.E.S.S. upper limits are shown for two spectral indices: 2.0 (red curve) and 3.0 (blue curve) from \cite{hess2014tev}. The green curve corresponds to synchrotron emission from regions with amplified magnetic fields, while the magenta curves represent synchrotron and IC on CMB emission fitted modeled for the lower limit field strength $B = 12~\mu \rm{G}$ as derived in \cite{hess2014tev}.}
    \label{fig:SED}
\end{figure}

We constructed a broadband spectral energy distribution (SED) spanning from radio to TeV gamma-rays. No significant gamma-ray signal was detected, and the resulting upper limit implies a lower limit on the magnetic field strength of $B \geq12~\mu \rm{G}$, assuming a leptonic scenario \citep{hess2014tev}. The SED shown in Figure \ref{fig:SED} indicates that the radio and X-ray emissions are produced by two distinct electron populations, as evidenced by the differing morphologies shown in Figure \ref{fig:region} (a). The green curve represents synchrotron emission from the first population, while the magenta curve corresponds to synchrotron and IC on CMB emission from the second population. This double electron population model was also used to reproduce the SED of SN 1006 \citep{tao2024observational}. In that case, the first population accounts for emission from compact regions or hot spots where the magnetic field is amplified to several mG. The second population emits in regions with an “average” magnetic field of several tens of $\mu$G. Since radio luminosity is proportional to the square of the magnetic field strength, radio emission from the hot spots is significantly enhanced. Thus, the radio emission is thought to be dominated by regions where the magnetic field is amplified by a factor of approximately 100. Similarly, the radio emission observed in G1.9+0.3 can also be interpreted as enhanced due to magnetic field amplification.

Magnetic field amplification is driven by turbulence generated through shock-cloud interactions, via the turbulent dynamo effect \citep{balsara2001evolution,  giacalone2007magnetic, inoue2011toward}.  \cite{inoue2011toward} demonstrated that the magnetic field in young SNRs can be amplified to $\sim 1 \rm{mG}$. Notably, the observed year-scale variability of synchrotron X-rays in RX J1713.7-3946 \citep{uchiyama2007extremely} can be explained by such shock-cloud interactions \citep{inoue2011toward}. When the primary shock collides with clouds, it generates reflection shocks that accelerate particles in turbulent regions where the magnetic field is amplified to around $1\rm{mG}$, producing short-time variability near the shocked clouds.

In G1.9+0.3, we propose that the shock in the northern region interacts with clouds, amplifying the magnetic field through the turbulent dynamo effect. Strong radio emission in this region may result from such amplification. If the magnetic field is amplified by a factor of 10-100, the corresponding radio luminosity could increase by a factor of 100 -10000. This is represented by the green curve in the SED. The magenta curves denote emission from regions with an “average” magnetic field, primarily contributing to the X-ray emission. The bipolar X-ray structure may reflect the orientation of the local magnetic field, as observed in SN 1006.

Our results for G1.9+0.3 reveal efficient particle acceleration (low value of $\eta$) in the radio rim where shock-cloud interaction occurs. In X-ray rims (polar region), the acceleration is less efficient, corresponding to a high value of $\eta$. Conversely, the acceleration is efficient in polar regions of SN1006 (e.g., \citeauthor{tsuji2021systematic} \citeyear{tsuji2021systematic}). We propose that this difference is caused by different mechanisms for generating turbulence between the two types of environments; the “clean” environments, such as the polar regions of SN 1006 and G1.9+0.3, and the environments where shock-cloud interaction occurs. The environments with shock-cloud interaction, such as RX J1713.7-3946 and radio rim of G1.9+0.3, the magnetic field is highly turbulent. This powerful turbulence would lead to the Bohm factor close to the Bohm limit ($\eta=1$). Indeed, \cite{tanaka2020shock} have reported that the particle acceleration of the southwestern part of RX J1713.7-3946 is close to the Bohm limit ($\eta =0.9^{+0.2}_{-0.3}$). In contrast, in “clean” environment, the turbulence is likely generated other processes, such as instability resulting from the pressure gradient derived from CR streaming (e.g., \citeauthor{bell1978acceleration}\citeyear{bell1978acceleration}; \citeauthor{caprioli2014simulations}\citeyear{caprioli2014simulations}). We believe that this mechanism generates a lower level of turbulence than that generated by shock-cloud interaction. For example, in SN 1006, the lowest Bohm factor ($\eta$) of the polar regions is $6.2 \pm 1.8$ \citep{tsuji2021systematic}.

Future high-resolution observations with the Atacama Large Millimeter/submillimeter Array (ALMA) could provide further insights into the shock-cloud interaction in G1.9+0.3. For instance, \cite{sano2020alma} identified 0.01 pc-scale cloudlets and filaments interacting with the northwest shell of RX J1713.7-3946 using ALMA. These features were located near X-ray hot spots exhibiting year-scale variability, which they attributed to mG-level magnetic field amplification resulting from shock-cloud interactions.

\section{Summary} \label{sec:summery}
We performed a spectral analysis of G1.9+0.3 using $Chandra$ observations, applying the ZA07 model to six distinct regions. In all regions, the spectral cutoff energy was found to be approximately 1 keV. Since this energy is proportional to the square of the shock velocity and inversely proportional to the Bohm factor $\eta$, we inferred that $\eta$ in the radio rim is nearly four times smaller than in the X-ray rims. This suggests that the magnetic field is highly turbulent in the radio rim, while it is less turbulent in the X-ray rims. Considering the decelerating shock and the presence of a CO cloud, the observed turbulence in the radio rim can be interpreted as the result of shock-cloud interaction. Such interactions are known to generate turbulence, which can amplify the magnetic field via the turbulent dynamo effect. The amplified magnetic field in the radio rim enhances the radio luminosity, likely forming the radio-bright structure seen in the northern region of G1.9+0.3. In contrast, the CO cloud located in the SW is suggested to lie in the foreground and not physically interact with the remnant. This is supported by the high Bohm factor and the high shock velocity in that region, estimated at $v_{\rm{sh}} =12,000-13,000 \rm{km~s^{-1}}$. Unlike the radio emission, the X-ray emission appears to originate from regions with an “average” magnetic field strength, where turbulence and amplification are less significant.

\begin{acknowledgments}
We thank the anonymous referee 
for his/her helpful comments to 
improve the manuscript.
This research was supported by Japan Science and Technology Agency (JST) ERATO Grant Number JPMJER21
02, Japan.
This paper employs a list of Chandra datasets, obtained by the Chandra X-ray Observatory, contained in~\dataset[doi:10.25574]{https://doi.org/10.25574/cdc.452}.
\end{acknowledgments}

\bibliography{G1.9+0.3_tao}{}

\begin{thebibliography}{}
\expandafter\ifx\csname natexlab\endcsname\relax\def\natexlab#1{#1}\fi
\providecommand{\url}[1]{\href{#1}{#1}}
\providecommand{\dodoi}[1]{doi:~\href{http://doi.org/#1}{\nolinkurl{#1}}}
\providecommand{\doeprint}[1]{\href{http://ascl.net/#1}{\nolinkurl{http://ascl.net/#1}}}
\providecommand{\doarXiv}[1]{\href{https://arxiv.org/abs/#1}{\nolinkurl{https://arxiv.org/abs/#1}}}

\bibitem[{Aharonian {et~al.}(2017)Aharonian, Sun, \& Yang}]{aharonian2017energy}
Aharonian, F., Sun, X.-n., \& Yang, R.-z. 2017, Astronomy \& Astrophysics, 603, A7

\bibitem[{Balsara {et~al.}(2001)Balsara, Benjamin, \& Cox}]{balsara2001evolution}
Balsara, D., Benjamin, R.~A., \& Cox, D.~P. 2001, The Astrophysical Journal, 563, 800

\bibitem[{Bell(1978)}]{bell1978acceleration}
Bell, A. 1978, Monthly Notices of the Royal Astronomical Society, 182, 147

\bibitem[{Borkowski {et~al.}(2017)Borkowski, Gwynne, Reynolds, Green, Hwang, Petre, \& Willett}]{borkowski2017asymmetric}
Borkowski, K.~J., Gwynne, P., Reynolds, S.~P., {et~al.} 2017, The Astrophysical Journal Letters, 837, L7

\bibitem[{Borkowski {et~al.}(2013)Borkowski, Reynolds, Hwang, Green, Petre, Krishnamurthy, \& Willett}]{borkowski2013supernova}
Borkowski, K.~J., Reynolds, S.~P., Hwang, U., {et~al.} 2013, The Astrophysical Journal Letters, 771, L9

\bibitem[{Brose {et~al.}(2019)Brose, Sushch, Pohl, Luken, Filipovi{\'c}, \& Lin}]{brose2019nonthermal}
Brose, R., Sushch, I., Pohl, M., {et~al.} 2019, Astronomy \& Astrophysics, 627, A166

\bibitem[{Caprioli \& Spitkovsky(2014)}]{caprioli2014simulations}
Caprioli, D., \& Spitkovsky, A. 2014, The Astrophysical Journal, 783, 91

\bibitem[{Carlton {et~al.}(2011)Carlton, Borkowski, Reynolds, Hwang, Petre, Green, Krishnamurthy, \& Willett}]{carlton2011expansion}
Carlton, A.~K., Borkowski, K.~J., Reynolds, S.~P., {et~al.} 2011, The Astrophysical Journal Letters, 737, L22

\bibitem[{Drury(1983)}]{drury1983introduction}
Drury, L.~O. 1983, Reports on Progress in Physics, 46, 973

\bibitem[{Enokiya {et~al.}(2023)Enokiya, Sano, Filipovi{\'c}, Alsaberi, Inoue, \& Oka}]{enokiya2023discovery}
Enokiya, R., Sano, H., Filipovi{\'c}, M.~D., {et~al.} 2023, Publications of the Astronomical Society of Japan, 75, 970

\bibitem[{Giacalone \& Jokipii(2007)}]{giacalone2007magnetic}
Giacalone, J., \& Jokipii, J.~R. 2007, The Astrophysical Journal, 663, L41

\bibitem[{Green {et~al.}(2008)Green, Reynolds, Borkowski, Hwang, Harrus, \& Petre}]{green2008radio}
Green, D., Reynolds, S.~P., Borkowski, K., {et~al.} 2008, Monthly Notices of the Royal Astronomical Society: Letters, 387, L54

\bibitem[{Grevesse \& Sauval(1998)}]{grevesse1998standard}
Grevesse, N., \& Sauval, A. 1998, Space Science Reviews, 85, 161

\bibitem[{{HESS Collaboration} {et~al.}(2014){HESS Collaboration}, Abramowski, Aharonian, Benkhali, Akhperjanian, Ang{\"u}ner, Anton, Balenderan, Balzer, Barnacka, {et~al.}}]{hess2014tev}
{HESS Collaboration}, Abramowski, A., Aharonian, F., {et~al.} 2014, Monthly Notices of the Royal Astronomical Society, 441, 790

\bibitem[{Inoue {et~al.}(2011)Inoue, Yamazaki, Inutsuka, \& Fukui}]{inoue2011toward}
Inoue, T., Yamazaki, R., Inutsuka, S.-i., \& Fukui, Y. 2011, The Astrophysical Journal, 744, 71

\bibitem[{Koo \& Heiles(1991)}]{koo1991survey}
Koo, B.-C., \& Heiles, C. 1991, Astrophysical Journal, Part 1 (ISSN 0004-637X), vol. 382, Nov. 20, 1991, p. 204-222., 382, 204

\bibitem[{Koo {et~al.}(1990)Koo, Reach, Heiles, Fesen, \& Shull}]{koo1990detection}
Koo, B.-C., Reach, W.~T., Heiles, C., Fesen, R.~A., \& Shull, J.~M. 1990, Astrophysical Journal, Part 1 (ISSN 0004-637X), vol. 364, Nov. 20, 1990, p. 178-186. Research supported by NSF and NASA., 364, 178

\bibitem[{Longair(1994)}]{Longair}
Longair, M.~S. 1994, {High Energy Astrophysics} (Cambridge University Press), \dodoi{10.1017/CBO9781139170505}

\bibitem[{Luken {et~al.}(2020)Luken, Filipovi{\'c}, Maxted, Kothes, Norris, Allison, Blackwell, Braiding, Brose, Burton, {et~al.}}]{luken2020radio}
Luken, K.~J., Filipovi{\'c}, M.~D., Maxted, N.~I., {et~al.} 2020, Monthly Notices of the Royal Astronomical Society, 492, 2606

\bibitem[{Reynolds {et~al.}(2008)Reynolds, Borkowski, Green, Hwang, Harrus, \& Petre}]{reynolds2008youngest}
Reynolds, S.~P., Borkowski, K.~J., Green, D.~A., {et~al.} 2008, The Astrophysical Journal, 680, L41

\bibitem[{Reynolds {et~al.}(2009)Reynolds, Borkowski, Green, Hwang, Harrus, \& Petre}]{reynolds2009x}
---. 2009, The Astrophysical Journal, 695, L149

\bibitem[{Sano {et~al.}(2020)Sano, Inoue, Tokuda, Tanaka, Yamazaki, Inutsuka, Aharonian, Rowell, Filipovi{\'c}, Yamane, {et~al.}}]{sano2020alma}
Sano, H., Inoue, T., Tokuda, K., {et~al.} 2020, The Astrophysical Journal Letters, 904, L24

\bibitem[{Tanaka {et~al.}(2020)Tanaka, Uchida, Sano, \& Tsuru}]{tanaka2020shock}
Tanaka, T., Uchida, H., Sano, H., \& Tsuru, T.~G. 2020, The Astrophysical Journal Letters, 900, L5

\bibitem[{Tao {et~al.}(2024)Tao, Kataoka, \& Tanaka}]{tao2024observational}
Tao, M., Kataoka, J., \& Tanaka, T. 2024, The Astrophysical Journal Letters, 970, L27

\bibitem[{Tsuji {et~al.}(2021)Tsuji, Uchiyama, Khangulyan, \& Aharonian}]{tsuji2021systematic}
Tsuji, N., Uchiyama, Y., Khangulyan, D., \& Aharonian, F. 2021, The Astrophysical Journal, 907, 117

\bibitem[{Uchiyama {et~al.}(2007)Uchiyama, Aharonian, Tanaka, Takahashi, \& Maeda}]{uchiyama2007extremely}
Uchiyama, Y., Aharonian, F.~A., Tanaka, T., Takahashi, T., \& Maeda, Y. 2007, Nature, 449, 576

\bibitem[{Zirakashvili \& Aharonian(2007)}]{zirakashvili2007analytical}
Zirakashvili, V.~N., \& Aharonian, F. 2007, Astronomy \& Astrophysics, 465, 695

\bibitem[{Zoglauer {et~al.}(2014)Zoglauer, Reynolds, An, Boggs, Christensen, Craig, Fryer, Grefenstette, Harrison, Hailey, {et~al.}}]{zoglauer2014hard}
Zoglauer, A., Reynolds, S.~P., An, H., {et~al.} 2014, The Astrophysical Journal, 798, 98

\end{thebibliography}
\bibliographystyle{aasjournal}

\end{document}